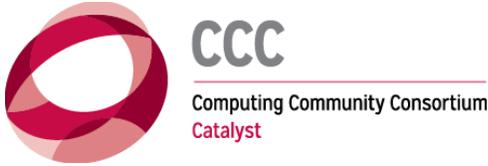

# Concerning the Responsible Use of AI in the US Criminal Justice System

Authors:

Cristopher Moore is a Professor at the Santa Fe Institute, Santa Fe, New Mexico, USA
Catherine Gill is a Program Associate and Government Affairs Liaison at the Computing Community Consortium, Washington, District of Columbia, USA
Nadya Bliss is Executive Director of the Global Security Initiative at Arizona State University, Tempe, Arizona, USA
Kevin Butler is a Professor and Director of the Florida Institute for Cybersecurity Research at the University of Florida, Gainesville, Florida, USA
Stephanie Forrest is Director of the Biodesign Center for Biocomputing and Professor at Arizona State University, Tempe, Arizona, USA
Daniel Lopresti is a Professor at Lehigh University, Bethlehem, Pennsylvania, USA
Mary Lou Maher is the Director of Research Community Initiatives at the Computing Community Consortium, Washington, District of Columbia, USA
Helena Mentis is a Professor and Department Head of Information Science at Drexel University, Philadelphia, Pennsylvania, USA
Shashi Shekhar is a Distinguished McKnight University Professor at University of Minnesota, Minneapolis, Minnesota, USA
Amanda Stent is the Director for the Davis Institute of AI at Colby College, Waterville, Maine, USA
Matthew Turk is the President of Toyota Technological Institute at Chicago, Chicago, Illinois, USA

Acknowledgements:

David Danks is a Professor of Data Science & Philosophy at the University of California, San Diego, San Diego, California, USA

Abstract:

Artificial intelligence (AI) is increasingly being adopted in most industries, and for applications such as note taking and checking grammar, there is typically not a cause for concern. However, when constitutional rights are involved, as in the justice system, transparency is paramount. While AI can assist in areas such as risk assessment and forensic evidence generation, its "black box" nature raises significant questions about how decisions are made and whether they can be contested. This paper explores the implications of AI in the justice system, emphasizing the need for transparency in AI decision-making processes to uphold constitutional rights and ensure procedural fairness. The piece advocates for clear explanations of AI's data, logic, and limitations, and calls for periodic audits to address bias and maintain accountability in AI systems.

*This opinion piece was written based on a response to a Request for Information issued by the National Institute of Justice. To read the full RFI response from the Computing Community Consortium, including independent members of the computing community, [please use this link](.).*

AI is advancing quickly and is being adopted in most industries. Using AI to draft an email or check your grammar is typically not a cause for concern, but using it to make decisions that affect people's lives is another matter. **When constitutional rights are involved, as in the justice system, transparency is paramount.**

During the Biden-Harris administration, Executive Order 14110 directed agencies to develop guidelines for acceptable uses and regulation of AI. Some of these uses, like summarizing and notetaking, will occur across the government. However, the nature and mission of each agency will create specific use cases for AI, such as monitoring threats at the DoD or simulating pollution at the EPA. The ways that agencies implement AI will change how they operate and may significantly impact the American people, and in some cases, people beyond our borders. One such area of application is the American criminal justice system.

The National Institute of Justice aims to "improve knowledge and understanding of crime and justice issues through science." They conduct research on controversial topics, such as predictive policing and risk assessment, to improve the justice system at all levels: from surveillance to forensic evidence to bail reform. In response to this Executive Order, the NIJ requested input on how AI will affect each layer of the justice system, and how to develop a plan to implement AI in ways that are accurate, fair, and constitutional. This article is a summary of our response.

That Executive Order has since been rescinded. Nevertheless, the current President's Executive Order 13859 calls for safe testing of AI and to "foster public trust and confidence in AI technologies and protect civil liberties, privacy, and American values in their application." How to protect these principles in the justice system, while using AI where appropriate, remains a vital question for us as computer scientists, for judges and policymakers, and for society as a whole.

Let's start by defining AI. We have yet to find a definition that doesn't cause many computer scientists to storm away in a huff; here we choose to define it broadly. "AI", as it is commonly understood, spans a wide range of technologies, from relatively simple algorithms that use statistical and machine learning techniques, to more advanced systems like deep neural networks and large language models. Some states have adopted the term "automated decision system" to include both current AI and simpler technologies, but we view these as belonging to one continuum.

AI has many obvious applications in the criminal justice system that can improve efficiency and effectiveness for members of law enforcement and the judiciary. AI can analyze criminal records, evidence, and legal documents much faster than humans and can identify patterns and connections that might otherwise be missed. Using AI to analyze legal documents and search for precedents can free up law enforcement officers and attorneys to focus on more strategic work. The danger that AI can perpetuate bias by being trained on biased data is well known; but at its best, AI may reduce bias by excluding factors correlated with race or income from its inputs, resulting in more equitable bail and sentencing decisions.

However, even when it satisfies statistical measures of accuracy and fairness, AI can pose significant concerns regarding *procedural fairness*, a vital aspect of the justice system. For one, AI is often not transparent. When an AI system makes a decision, it is not always clear — to defendants, judges, or other stakeholders — how the system came to its conclusions. This is especially concerning in bail, sentencing, and parole. Denying a citizen their liberty is one of the most fundamental and momentous actions a government can take. If this decision is supported in part by an AI system, then that citizen (and their defense counsel) need to know what data were used by the AI, where the data came from, and the logic by which its recommendation was produced from this data.

The possibility that an opaque AI — which neither defendants, nor their attorneys, nor their judges understand — could play a role in major decisions about a person's liberty is repugnant to our individualized justice system. **An opaque system is an accuser the defendant cannot face; a witness they cannot cross-examine, presenting evidence they cannot contest.**[8] In our view, any AI system used for criminal justice must be transparent, as opposed to being a "black box" that produces outputs using a hidden process.

By "transparent" we do not necessarily mean making an AI system "open source," i.e., publishing the source code of its program. While this may be necessary in some domains, in general it is neither sufficient nor necessary. Transparency means a clear description of the internal workings of an AI system: the mathematical and logical methods it uses to produce its output. A system should be transparent enough so that its logic is intelligible to the decision makers it advises and the people it affects, and so that it can be independently tested to see if it performs with the accuracy and characteristics claimed by its developers.

Some tech companies complain that requiring this kind of transparency would violate their intellectual property rights, reveal trade secrets, or discourage innovation. We reply that opaque, proprietary AI systems might be acceptable in certain domains — recommending movies, translating speech, and so on — but they should not play a role in the justice system of a society that values individual rights and government accountability. If the government is considering the use of an AI system, the public has every right to require transparency, and this requirement should be implemented in procurement policies. The public might also demand an explanation, not just of what input factors and weights the AI system uses, but why it uses those factors and those weights: in technical terms, the feature selection and training process.

Beyond providing an explanation of an AI's decision, attorneys for the defense and prosecution, as well as judges, need to possess a fundamental understanding of how these systems operate. This knowledge is crucial because it enables attorneys to identify potential biases or errors in the AI's decision-making process and formulate cogent arguments for or against its findings. Similarly, judges can't hope to fairly and accurately weigh an AI's recommendations if they have no understanding of how it generates them. We recommend a repository of educational resources so that attorneys and judges can learn what data AIs are being trained on, how these data are being collected, and how these systems arrive at their recommendations. This understanding is vital for all parties to properly use AI, contest it, and trust it when trust is warranted.

When relying on AI systems to make judicial recommendations, such as in sentencing or probation decisions, judges should always view these systems' results through a critical lens. AI systems are often subject to biases and operate based on limited and/or partial data. An AI system typically uses prior convictions and prior failures to appear when recommending a bail decision. It treats each defendant as a member of a group, namely those with similar criminal records. But it has no access to individualized facts about a defendant that might make them more, or less, dangerous than others in this group. The judge should be open to additional arguments made by the defense and prosecution, whose job it is to present information not considered by the AI.

Transparency also applies to the decision-making process: not just the AI itself, but how humans use it. How and when AI systems are used in courtrooms must be standardized and fully explained, and all parties involved in a particular case must be informed when the results of an AI contribute to a decision. Standardizing procedures regarding when AI can and should be used will also improve the processes of auditing and evaluating the use of AI in the criminal justice system. Demanding transparency from AI can make the justice system as a whole more transparent.

AI systems — and the social science and data science behind them — can also help advance policy discussions about crime and justice. But to do this, risk assessment systems should not lump all types and levels of crime together.[6] Unfortunately, while they typically distinguish violent from non-violent charges, many risk assessments in use today do not make a distinction between felony and misdemeanor arrests, even though many legal scholars have pointed out the need to do so.[9] Of course, a judge might feel that the risk of a misdemeanor is enough to justify pretrial detention. But if an AI lumps all kinds of risk together, it does not help the judge consider the "nature and seriousness of the danger to any person or the community that would be posed by the person's release" as the law requires.[10]

Worse, many risk assessments provide the judge with an abstract label like "high risk," as opposed to a quantitative estimate of the probability of rearrest. This points out the need for another kind of transparency: judges should know what an AI's output actually means. Phrases like "high risk" allow our preconceptions and stereotypes to run wild. Psychologists have found that human decision-makers often overestimate the probability of bad events: mock jurors given categorical labels like "high risk" greatly overestimate the corresponding probabilities.[4] To be useful and accountable, an AI system should make predictions that are as specific and quantitative as possible: it should say how much risk, and risk of what.

Moreover, the meaning of an AI's output might not translate from one jurisdiction to another.[6] Due to differences in demographics, policing, and many other factors, an AI might display racial bias in one state or city even though it is unbiased in another. The performance of AI systems can also change over time: for instance, new diversion programs might reduce the level of risk some defendants pose. In that case, using a risk assessment trained on data before those programs were implemented will overestimate risk.[3] For this reason, it is vital to perform periodic audits of risk assessment algorithms in each jurisdiction they are used, whenever there is enough data to obtain good statistics on their accuracy and various measures of bias.[7] California currently requires that pretrial risk assessment algorithms be audited every three years.[1] Other states should follow suit and extend similar requirements to AIs used at every layer of the justice system including sentencing, parole, prison classification, and forensic evidence.

Finally, an AI system should provide information about its uncertainty and the confidence level of its predictions. If a defendant has an unusual criminal record, with very few similar defendants in the training data, an AI should report that its output is less certain. If an AI model generates a prediction or determination that is highly unusual or that humans struggle to justify, those results should not be considered by the court as evidence in support of or in opposition to a defendant. Unusual edge cases where the AI model makes confusing predictions should also be reported to the developers to improve the model in the future. Developing principled techniques for calculating confidence levels is an active area of research, as is developing effective ways to communicate uncertainty to human decision-makers.

While we focused here on risk assessments, many of these principles apply just as well to the use of AI to generate forensic evidence, such as facial recognition and probabilistic genotyping of DNA samples. Despite the weight given to this kind of evidence by judges and juries, many of these software products have never been independently validated. Developers typically oppose revealing anything about their inner workings unless forced to do so by a judge, again threatening the right of defendants to confront evidence against them and cross-examine witnesses.[11] Indeed, systems that have been used for DNA evidence in thousands of cases have turned out to have important flaws when independent studies are finally carried out.[5] We agree that judges and legislatures should require independent verification and validation of these systems, just as we do for software in other high-risk areas like medicine or power plant management.[2]

Where does this leave us? AI can help us make consequential decisions, and can make human decision-making more accountable and fair. But only if it is transparent, so that those affected by it — and the decision-makers advised by it — understand what data the AI uses, what it does with this data, and what mistakes it can make.

The law is a dynamic system. It has been in development for thousands of years, since the first humans disagreed and found resolution. A key pillar of a just legal system is a human-centered design where evidence and arguments can be contested and disputed, and where the system itself can be revised to make its outcomes and processes more just. In our society we believe this system should be accountable, fair, and transparent, and open to the unique characteristics of individuals. We should not throw these values away in pursuit of technological improvements.